\documentclass[prd,aps,preprint,tightenlines,nofootinbib]{revtex4}

\begin{document} 

\title{Anomalous Observers in the Subjectively Identical Reference Class }
\author{Mike D. Schneider}
\author{Ken D. Olum}
\affiliation{Institute of Cosmology,
Department of Physics and Astronomy, Tufts University, Medford MA 02155, USA}

\begin{abstract} 

Anthropic reasoning is a critical tool to understand probabilities,
especially in a large universe or multiverse. According to anthropic
reasoning, we should consider ourselves typical among members of a
reference class that must include all subjectively indistinguishable
observers.  We discuss here whether such a reference class, which we
assume must include computer simulations, must also include computers
that replay previous simulations, magnetic tapes that store but do not
``run'' the simulation, and even abstract mathematical functions.  We
do not see any clear criterion for excluding these anomalous observers,
but their presence is deeply troubling to the idea of anthropic
reasoning.

\end{abstract}

\maketitle

\section{Introduction}

Suppose the universe is large enough that there are instances of
multiple intelligent life forms, or even multiple Earths, spread out
at distances too vast to allow any interaction. In such a universe,
different populations will observe different things happening around
them. If the universe is large enough, then any outcome will likely be
observed by one of these populations.

In the face of such a situation, how can we say that one thing is more
likely then another?  Anthropic reasoning is a tool used to understand
probability in such a universe. Vilenkin's \cite{gr-qc/9406010}
Principle of Mediocrity says that we should ``think of ourselves as a
civilization randomly picked in the metauniverse.'' More specifically,
Bostrom's \cite{Bostrom:book} Self-Sampling Assumption (SSA)
says that ``One should reason as if one were a random sample from the
set of all observers in one's reference class.''  By ``observer'' we
mean any conscious agent.\footnote{The term ``observer'' for this role
  may be an odd choice; perhaps ``thinker'' would be better. Anthropic
  ideas were originally discussed for observation of the ``constants''
  of nature, which might vary across the universe. Since these
  origins, the use of ``observer'' has stuck.} The \emph{reference
  class} is the class of all observers who are sufficiently alike that
we could be, or could have been, any of them.

If the reference class is well defined, then the anthropic idea of
typicality has predictive value.  Consider the following example.
Physicists working at CERN find a particle behaving like the
theoretical Higgs Boson with a mass-energy of 125 GeV.  They state a
confidence of five sigma, meaning that the chance their observations
would have occurred in the absence of a Higgs is less than one in a
million, and thus they can be almost certain that they have discovered
the Higgs.  But now suppose the Higgs does not exist, and also suppose
that we are within a multiverse in which there are many millions of
copies of these physicists performing the same experiment looking for
the Higgs.  For simplicity, assume that the multiverse is very large
but finite.  We can expect that one in a million of these
physicist-copies would find the Higgs, even though it is not there.
Philosophers here on this Earth might now say ``Perhaps our physicists
are those who unluckily have discovered evidence for the Higgs, even
though it does not actually exist.  Who knows?''

The anthropic resolution to this concern answers simply that the
physicists must consider their universe to be \emph{typical} among all
the universes in the finite multiverse.  Thus, if erroneous data appears
by chance only in one out of a million universes, the chance that it
has appeared here is only one in a million. Therefore, we can be
quite confident of having correctly discovered the Higgs.

The precise definition of the reference class has been a matter of
considerable debate.  But, for the example above, it seems sufficient
to consider only the class of subjectively indistinguishable
observers.  Each observer in that class cannot distinguish himself
from any other observer in the class: they have the same memories and
are having the same experiences and thoughts.  Thus it seems that any
reference class must be at least as large as the class of subjectively
indistinguishable observers, since by definition one could be any of
those.

Included in this idea is the possibility that two observers can be in
two different settings, but while they are unable to determine which
of those settings either one resides, they must be in the subjectively
identical reference class.\footnote{In its strongest form, anthropic
  reasoning imagines that we know all information about the universe
  (including the number of duplicate observers), except where we are
  located within that universe. This premise is familiar from the
  literature on self-locating beliefs.  Bostrom
  \protect\cite{Bostrom:belief} distinguishes anthropics from
  philosophy of self-locating beliefs: anthropics is a method to
  account for selection effects \emph{in the face of self-locating
    doubt}.}

Larger reference classes are often used, but that will not be our
concern here.  Instead we will argue that even the class of
subjectively indistinguishable observers may be distressingly large,
containing exotic observers we would rather have left out.  This
paper should be read as a descent into increasing discomfort: we
demonstrate first that the somewhat uncontroversial inclusion of
simulations in the reference class (Sec.~\ref{sec:computers}) carries
with it ``slave computers'' (Sec.~\ref{sec:slaves}), magnetic tapes
and similar static data storage devices (Sec.~\ref{sec:tapes}),
algorithmically compressed data (Sec.~\ref{sec:compression}), and
eventually even mathematical functions
(Sec.~\ref{sec:mathematics}). We conclude in Sec.~\ref{sec:conclusion}
by discussing the status of anthropic reasoning in the face of these
anomalous observers.


\section{Computer Simulations}
\label{sec:computers}

We first consider computer simulations of humans.  A sufficiently
large computer could simulate all the atoms in a human body, or it
could simulate the operation of the neurons that make up the brain, or
perhaps even simulate human thought directly.  Bostrom
\cite{Bostrom:computersimulations} argues that we will be able
to build computers capable of simulating humans in the not-too-distant
future.  We might expect some of the simulations to be intentionally
simulated copies of you or me, and in a sufficiently large universe we
should expect some of the simulations coincidentally to be copies of
you or me.  We assume here that human consciousness arises from the
operation of the biochemical constituents of the human body and brain,
and therefore that, at a minimum, faithful simulation of these
constituents would lead to a faithful simulation of human
consciousness.

Since there is a one-to-one correspondence between the states of the
simulated brain and the states of a real brain, the observer in such a
case cannot distinguish which of the two she is.  Thus, it follows
simply that simulations should be included in the reference class of
subjectively identical observers.

A computer simulation is somewhat analogous to the ``brain in a vat''
that has occasioned much discussion.  Advocates of Putnam's response
to brain-in-a-vat skepticism may try to argue a similar case to
dismiss simulations from the reference class.  The argument would go
as follows.  Suppose in fact I am a computer simulation, and suppose I
say ``I am a computer simulation''.  When I say ``computer
simulation'', these words refer not to the type of thing I actually
am, but to another type of thing that might be performed on a
``simulated computer'' in my simulated world.  Therefore, whether or
not I am in a simulation, the statement ``I am a computer simulation''
is false when I say it.  By this, the argument goes, I cannot be a
computer simulation.

We will give two responses to this claim.  For a moment, let us accept
the externalist viewpoint.  Thus, we will concede that when the human
wonders whether she is a computer simulation, she is wondering about
something different than what the simulation wonders in its equivalent
processes.  If these entities could distinguish between the two
possibilities they could wonder about, then they would not be
subjectively identical.  But it is only the external differences that
lead to the difference in the content of their thoughts; these
differences are not accessible to them.  Thus even though their
thoughts are, on the externalist reading, different, they do not have
access to the externals that distinguish these thoughts, and so are
still subjectively indistinguishable.

Our second response to Putnam's sort of reasoning is that ``computer
simulation'' is a generic term analogous to ``envatted''
\cite{DChalmers2005} rather than a natural kind term like ``brain''.
In particular, even if ``sims'' have access only to sim computers, the
idea of a computer and its use is universal (the same for sims and for
organic humans), because it is essentially a mathematical concept, as
formalized by Turing \cite{Turing:1936}.  By this analysis, sims and organic
humans mean the same thing when they utter ``I am a computer
simulation'', and Putnam's argument does not apply.  We can also
imagine an organic mind whose brain state will be copied into a computer
simulation.  Presumably, at least at first, the simulation's thoughts
and utterances have the same referents that they had when the thoughts
lived in the original organic mind (see \cite[p. 81]{Wright:Putnam}
and the ``recent matrix hypothesis'' of \cite{DChalmers2005}).  These
are by no means exhaustive accounts of the philosophical foundations
for simulations, but we feel that they are sufficient to preserve the
feasibility of simulations in the anthropic reference class of
subjective indistinguishability.

To understand how simulations could arise in anthropic questions,
suppose that when you go to sleep tonight, the state of your brain is
copied into a computer simulation, and that you know that this is
going to happen.  If this simulation is part of the reference class,
when you awake tomorrow, you should think the odds are 1:1 that you
are now the computer simulation.  If there are 100 computer
simulations, you should think that the odds are 100:1.

In the latter case, we have assumed that duplicate computer
simulations count as separate observers in the reference class.  We
will now explore whether this is a safe assumption to make.  One issue
arises because of the use of redundancy in computer design.  Consider
a computer that, for the sake of reliability, completes every process
and computation twice.\footnote{Such computers have in fact been
  built, for example, by Tandem Computers, Inc.}  If we want to count
all duplicate simulation programs as separate, it appears we must
consider such a system as two observers.  But why should it matter to
the philosopher how the computer engineer did her job?\footnote{For
  that matter, consider the low-level design of the computer.  Every
  operation in an electronic computer is represented by the movement
  of a large number of electrons.  So, in a certain sense, all
  computers have a high degree of redundancy.}  Anthropic reasoning
has given us a window, which probably should not exist, into the
precise design of the computer running a simulation.

So what of the alternative: identical simulation programs count
collectively as one observer?  Again consider 100 identical computer
simulations.  They all count as one observer.  But now imagine several
of the computers malfunction and the programs
deviate \cite{Dennett:WhereAmI}.  Do these deviating simulations
suddenly count as multiple observers in the reference class? It is
easy to imagine multiple observers if the deviations are significant,
but consider 100 simulations at the atomic level: do we count multiple
observers if the deviations are as insignificant as a single atom?
This is a weak reason for the spontaneous creation of a new observer:
the change is indistinguishable.  There does not seem to be any
non-arbitrary cutoff that justifies a new observer.

Another problem is that just by changing a tiny detail of the
simulation that does not enter into the simulated observer's
consciousness, we can split or combine observers.  Suppose there are
10 groups of 100 simulations each.  The simulations in each group are
identical, but the groups are different from each other because they
have decided on different bets in a 10-horse race.  Now suppose that
after the race is run, the simulation program makes a tiny change to
each of the 100 simulations that bet on the winning horse.  Those 100
simulations thus become different observers, and there are 109
observers: 100 winners and 9 losers.  Thus, before the race is run,
each observer thinks her chance to win is 1/10, but afterwards
(perhaps even before the result is posted), she thinks the chance that she has won is 100/109, quite an anomalous result.

We might avoid this type of problem by saying that all deviating
observers count as separate observers in the reference class at any
point in time, even before the observers deviate.  But this is even
worse, because it allows communication backward in time. If we apply
this analysis to the above example, we find that the 109 different
observers exist even before the race is run.  Thus immediately upon
placing her bet, every observer should believe with confidence 100/109
that she will be a winner.

The above reasons lead us to reject the proposal that only deviating
observers count separately, and we return to the assumption that
duplicate observers count separately in the reference class in all
circumstances.\footnote{An organic brain is even more redundant than
  an electronic computer, and one might ask whether we should now
  count one human as a large number of duplicate observers in the
  reference class.  This seems quite an unattractive idea, but it
  might have the interesting effect of providing a solution to the
  ``doomsday argument'' \protect\cite{Carter:original,Leslie:1989}.
  Perhaps today's highly redundant biological observers will be
  succeeded by many more observers running on very efficient
  computers, but because each of us biological humans appears many
  times in the reference class, we might still be typical among all
  observers to ever exist.}


\section{Slave Computers}
\label{sec:slaves}

So far, we have not said anything about the operation of a computer
simulation. We will use the following simple model.  The computer has
a large memory which stores a representation of the current state of
the brain being simulated.  Data structures in the memory could
represent thoughts, neurons, or atoms, depending on the level of the
simulation.  There is then a program that advances the simulation to
the next state, based on the previous state and inputs representing
the sensory information going into the brain.  This program runs over
and over again, advancing through a sequence of states representing
the states of a thinking brain. The interval of simulated time between
one state and the next should be small compared to any timescale of
the process being simulated (e.g., the timescale of chemical reactions
in atomic-level simulation).  This allows a discrete-time simulation
to accurately model a continuous physical process.  Simulation
techniques such as this are widely used in science.

Suppose there is such a computer simulation running, and now consider
a second computer that works in a different way.  It has the same type
of memory and processor as the first computer, but rather than
deriving each state from the next, this second computer merely copies
each state that the first computer generates.  This ``slave computer''
thus goes through exactly the same states as the principal computer
whose memory it is copying.

Does such a slave computer count as a separate member of the reference
class?  Suppose, for example, you know that you exist as a computer
simulation that is being run on 2 principal computers A and B.
Suppose right now the simulations are identical, but soon simulations
A and B will have different experiences, $F_A$ and $F_B$.  Then, your
credence that you will soon experience $F_A$ should be 50\%.  But now
suppose A has 99 slaves while B has none.  Should you now consider the
odds to be 100:1 that you will experience $F_A$ rather than $F_B$?

We might try to exclude slave computers from the reference class by an
appeal to causal relationships.  The successive states of the slave
computer are not causally related to each other, but rather each is
passively copied from the principal computer.  Perhaps this does not
constitute ``thinking'', and it would not be possible for one to be such
a slave computer.

Unfortunately, there is a problematic feature in this appeal to
causality.  Suppose the second computer at each step first copies the
state from the principal computer, but then computes this same state
by operation of the regular program (on a saved copy of the previous
state).  Does this second step make the computer not a slave, because
the ``direct cause'' of the state is computation rather than copying?
The second step consists of storing some data into memory, when that
exact same data was in that memory already, so it seems rather
vacuous.  Suppose the second computer does the computation, but then
discards the result, which is hardly different from storing it in the
locations that contain it already.  Would it now be excluded from the
reference class?

If copy followed by computation admits the second computer to the
reference class, then what about the reverse case, where the state is first
computed and then overwritten with an identical state produced by
copying?  In this case, the ``direct cause'' is the copying of the data
(into locations that already contain it).  In this case, is the second
computer now to be excluded from the reference class?

These considerations appear to weaken the intuition that there exists
a distinction based on causality, and so we must be open to the idea
of including slave computers in the reference class.


\section{Magnetic Tapes}
\label{sec:tapes}

Now consider a slightly different situation.  The principal computer
runs the simulation, computes the successive brain states, and writes
them onto magnetic tape.  At a later time, the slave computer copies
the states one by one from the tape into its working memory.  On the
grounds that the details by which the data travel from one computer to
another should not matter, it seems that this slave computer also
should belong in the reference class.

But what is the importance of actually reading the tape?  Suppose that
the memory of the slave computer is broken, so that it cannot be read.
This should not make a difference, because this computer never reads
its memory.  Now suppose instead that the memory is broken so that the
data written to it is never actually stored.  Does this exclude the
computer from the reference class?  Why should it matter whether the
data was successfully stored in the memory when it cannot be read?
Functionally, there is no difference between unreadable and unwritable
memory, so why should it matter here?

In a real computer, it is possible to page through a file by keeping
the entire file in physical memory and merely adjusting pointers so
that successive parts of the file appear in the same location in the
virtual memory of a computer program.  This operation is equivalent to
reading successive sections of the file into memory.  Does the program
that does nothing but manipulate pointers count as a member of the
reference class?

Perhaps the reading of the tape does not matter, and the tape itself
is a member of the reference class.  But this seems quite
counterintuitive; should I really be concerned that I am just some
static data sitting somewhere on a tape?\footnote{The magnetic tape
  can be considered a realization of the eternalist idea that
  separation in time is analogous to separation in space.  The
  different moments in the observer's consciousness are physically
  positioned in different places along the tape.  If one considers
  eternalism as viable, then magnetic tape observers are not as
  far-fetched as they may at first seem.} If the tape is duplicated,
does the existence of the second tape constitute a second observer?
Each bit on a magnetic tape is stored in the spins of many electrons
in a region of the magnetic material.  Does this redundancy mean that
each tape itself constitutes many observers?


\section{Compressed data}
\label{sec:compression}

Presumably, the precise way in which the successive brain states are
written on the magnetic tape does not determine whether the tape
belongs in the reference class. A tape drive today might well have a
data compression algorithm, so that what appears on the tape is not
the actual data, but a compressed version of it that in common cases
will use less space on the tape.  The existence of such a system
should be of no consequence for membership in the reference class.

But note that the program which produces each brain state from its
predecessor is just evaluating some mathematical function, so, say,
$x_n = f(x_{n-1})$ where $x_n$ is the current state and $x_{n-1}$ is
its predecessor.  Thus $x_n$ can be found by applying $f$ successively
$n$ times to some initial state $f_0$.  This fact permits an
extremely efficient compression procedure, in which rather than
writing $x_n$ on the tape, we just write $n$, and the reading program
understands this to mean $x_n$.  But then, if the tape containing
the normally compressed brain states is to be considered an observer,
what about the tape which just contains successive integers that
constitute super-compressed versions of the brain states?  To take
this absurd conclusion even further, given any set of data there is
some decompression algorithm that turns it into any other set of
data.  Does this mean that any set of data constitutes an observer
subjectively indistinguishable from me?


\section{Pure mathematics}
\label{sec:mathematics}

The successive states of the computer simulation, $x_n$ are just large
collections of bits, which could equally well be interpreted as
integers.  But integers, or a sequences of integers, exist as purely
mathematical objects whether or not some actual computer happens to be
computing those numbers.  So, should we think that the observer
represented by the sequence $\{x_n\}$ or, alternatively, by the number
$x_0$ and the function $f$, exists as an object in the abstract world
of mathematics?  That is to say, could we be confused as to whether we
are mathematical abstractions \cite{Tegmark:multiverse}?

If our lives up to the present are represented by a finite sequence of
integers, and different future possibilities are represented by
different continuations of this sequence, what does it mean for one
future to be more likely to another?  It doesn't make sense to say
that a given sequence can exist more than once as an abstract
mathematical entity, but if all sequences exist just once it seems that
that all possible futures are equally likely, which would make
nonsense of our normal ideas of probability.


\section{Conclusion}
\label{sec:conclusion}

In a large universe, there might be many people identical to us.  They
may have the same history that we have, but their futures may be
different.  Since we don't know which of these people we are, we make
predictions by considering ourselves as randomly chosen members of
some reference class.  It seems that this reference class must contain
all observers subjectively identical to us, because any such observer
cannot distinguish herself from any other such
observer.\footnote{There is a parallel between subjectively
  indistinguishable observers excluded from the reference class and
  \emph{philosophical zombies}, i.e., entities who behave in every way
  as conscious beings would but are nevertheless not conscious, so
  that one never has to consider the possibility that one might be one
  of the zombies.  All arguments against the possibility of zombies,
  e.g., \protect\cite{Dennett:Zombies} militate against the exclusion of any
  subjectively indistinguishable observer from the reference class.}

Unfortunately, it appears that even the class of subjectively
indistinguishable observers is much larger than one would want.
Computer simulations of human beings should be included, but then it
becomes hard to draw a line between such simulations and other cases
that are quite anomalous.  We have argued that once computer
simulations are included, one should also include computers that do
nothing but copy ``brain states'' from one place to another, or even
that the list of brain states without any action should be considered
an observer.  If that is true, maybe the list of numbers representing
states, viewed as a mathematical abstraction, should also be included.

Including such anomalous members of the reference class seems
ridiculous, so perhaps anthropic reasoning should be abandoned.  But
then it is hard to make sense of everyday ideas about the likelihood
of future events.  How can I give a probability distribution to what
may happen to me tomorrow, if some copy of me will experience every
possible future?  The basic idea of anthropic reasoning seems
necessary, and it adequately solves the simplest case of a large,
finite universe with a finite number of duplicate observers, so we
would be reluctant to give it up.

The underlying problem here seems to be that the results of anthropic
reasoning may depend on the answers to deep questions in the
philosophy of mind and identity theory.  While we would like to know
the answers to such questions as ``Which entities are possibly us?''
and ``When are two copies the same person and when are they two
people?'', these answers traditionally have no practical effect.
Somehow, anthropic reasoning seems to privilege these philosophical
queries by granting them influence on our reasoning about which
theories of physics and cosmology are correct.

We feel this poses a serious puzzle.  In the absence of any
alternative theory to make sense of probabilities, there should be a
way to utilize anthropic ideas without including anomalous copies of
ourselves.  Unfortunately at present we are not able to find any
solution that we feel is adequate.

\section*{Acknowledgments}

We would like to thank Adam Brown and Daniel Dennett for helpful
conversations.  This research was supported in part by grant
MGA-08-006 from The Foundational Questions Institute (fqxi.org)

\bibliographystyle{unsrt}
\bibliography{anthropics}

\end{document}